\documentclass[aps,prl,floats,floatfix,nofootinbib,preprintnumbers]{revtex4-1}
\usepackage{graphicx,amsmath}
\usepackage{amssymb}
\usepackage{amsfonts}
\usepackage{hyperref}
\usepackage[bottom]{footmisc}
\usepackage{subfigure}

\newcommand{\laem}{\stackrel{<}{\sim}}

\begin{document}

\preprint{MSUHEP-110819}
\preprint{FERMILAB-PUB-11-386-T}

\title{LHC Limits on the Top-Higgs in Models with Strong Top-Quark Dynamics}

\author{R.\ Sekhar Chivukula}
\email {sekhar@msu.edu}
\affiliation {Department of Physics and Astronomy,
Michigan State University,
East Lansing, MI 48824, USA}

\author{Baradhwaj Coleppa}
\email {barath@physics.carleton.ca }
\affiliation {Ottawa-Carleton Institute for Physics,
Carleton University,
Ottawa, Ontario K1S 5B6, Canada}

\author{Heather E.\ Logan}
\email {logan@physics.carleton.ca }
\affiliation {Ottawa-Carleton Institute for Physics,
Carleton University,
Ottawa, Ontario K1S 5B6, Canada}

\author{Adam Martin}
\email {aomartin@fnal.gov }
\affiliation {Theoretical Physics Department,
Fermilab, 
Batavia, IL 60510, USA}

\author{Elizabeth H.\ Simmons}
\email {esimmons@pa.msu.edu}
\affiliation {Department of Physics and Astronomy,
Michigan State University,
East Lansing, MI 48824, USA}

\date{\today}

\begin{abstract}
LHC searches for the standard model Higgs Boson in $WW$ or $ZZ$  decay modes
place strong constraints on the top-Higgs state predicted in many models with new dynamics preferentially affecting top quarks.
Such a state couples strongly to top-quarks, and is therefore produced through gluon fusion at a rate enhanced
relative to the rate for the standard model Higgs boson. 
A top-Higgs state with mass less than 300 GeV is excluded at 95\% CL if the associated top-pion has a mass of 150 GeV,  
and the constraint is even stronger if the mass of the top-pion state exceeds the top-quark mass or
if the top-pion decay constant is a substantial fraction of the weak scale.  These results have significant implications for 
theories with strong top dynamics, such as topcolor-assisted technicolor, top-seesaw models, and certain Higgsless models.

\end{abstract}

\maketitle

\section{Introduction}

The primary mission of the Large Hadron Collider is to uncover the agent of electroweak symmetry breaking and
thereby discover the origin of the masses of the elementary particles. In the standard model \cite{standard-model},
electroweak symmetry breaking occurs through the vacuum expectation value of a fundamental weak-doublet
scalar boson. Via the Higgs mechanism \cite{higgs-mechanism}, three of the scalar degrees of freedom of
this particle become the longitudinal states of the electroweak $W^\pm$ and $Z$ bosons and the last, the standard
model Higgs boson ($H_{SM}$), remains in the spectrum. Recently, both the ATLAS \cite{ATLAS-higgs} and CMS \cite{CMS-higgs} collaborations have set limits on the existence of a standard model Higgs boson.
In this paper we apply these limits to the ``top-Higgs" ($H_t$) expected in topcolor assisted technicolor models and other models with new strong dynamics preferentially affecting the top quark.

Topcolor assisted technicolor (TC2) \cite{Hill:1994hp,Lane and Eichten - TC2,Popovic:1998vb,Braam:2007pm} is a dynamical theory of electroweak symmetry breaking that combines the ingredients of technicolor \cite{Weinberg:1979bn,Susskind:1978ms} and top condensation \cite{Miransky:1988xi,Miransky:1989ds,Nambu:1989jt,Marciano:1989xd,Bardeen:1989ds}. Top condensation and the top quark mass arise predominantly from ``topcolor" 
\cite{Hill:1996te}, a new QCD-like interaction that couples strongly to the third generation of quarks.\footnote{Additional interactions are also included to prevent formation of a $b$-quark condensate and, hence, allow for a relatively light $b$-quark; the simplest example \cite{Hill:1994hp} is an extra $U(1)$ interaction, giving rise to a topcolor $Z'$; other ideas are discussed in \cite{Buchalla:1995dp}.} 
Technicolor then provides the bulk of electroweak
symmetry breaking via the vacuum expectation value of a technifermion bilinear.

TC2 is an important potential ingredient in theories of dynamical electroweak symmetry breaking \cite{Hill:2002ap}. 
In particular, it is difficult to construct technicolor theories which accommodate the heavy top-quark
without also producing large and experimentally forbidden corrections to the ratio of 
$W$- and $Z$-boson masses \cite{Chivukula:1988qr} or to the coupling of the $Z$-boson
to bottom-quarks \cite{Chivukula:1992ap}. By separating the sector responsible for top-quark mass generation from that
responsible for the bulk of electroweak symmetry breaking, TC2  alleviates these difficulties. The ``top-triangle moose" model \cite{Chivukula:2009ck}, combining Higgsless and topcolor models, is a consistent low-energy effective theory for models with separate sectors for generating the top mass and the vector boson masses.   It can be used to investigate the phenomenology of TC2 theories \cite{Chivukula:2011ag} and other theories with strong top dynamics, and we employ it in that capacity in this analysis.

As we review below, theories with strong top dynamics 
generically include a top-Higgs state -- a state with the same quantum numbers as the standard model
Higgs boson, a mass of generally less than 350 GeV, and a stronger coupling to top-quarks than
the standard model Higgs has. We show that the ATLAS \cite{ATLAS-higgs} and CMS \cite{CMS-higgs} searches 
for the standard model Higgs exclude, at 95\% CL, a top-Higgs with a mass less than 300 GeV provided that the associated top-pion states have a mass of at least 150 GeV; even  heavier top-Higgses are also excluded by the data under certain conditions.  These results constrain model-building in theories with strong top dynamics.

\section{Top-Color Assisted Technicolor: Scalar Spectrum and Properties}

\begin{figure*}[ht]
\begin{center}
\includegraphics[angle=0,width=3.5in]{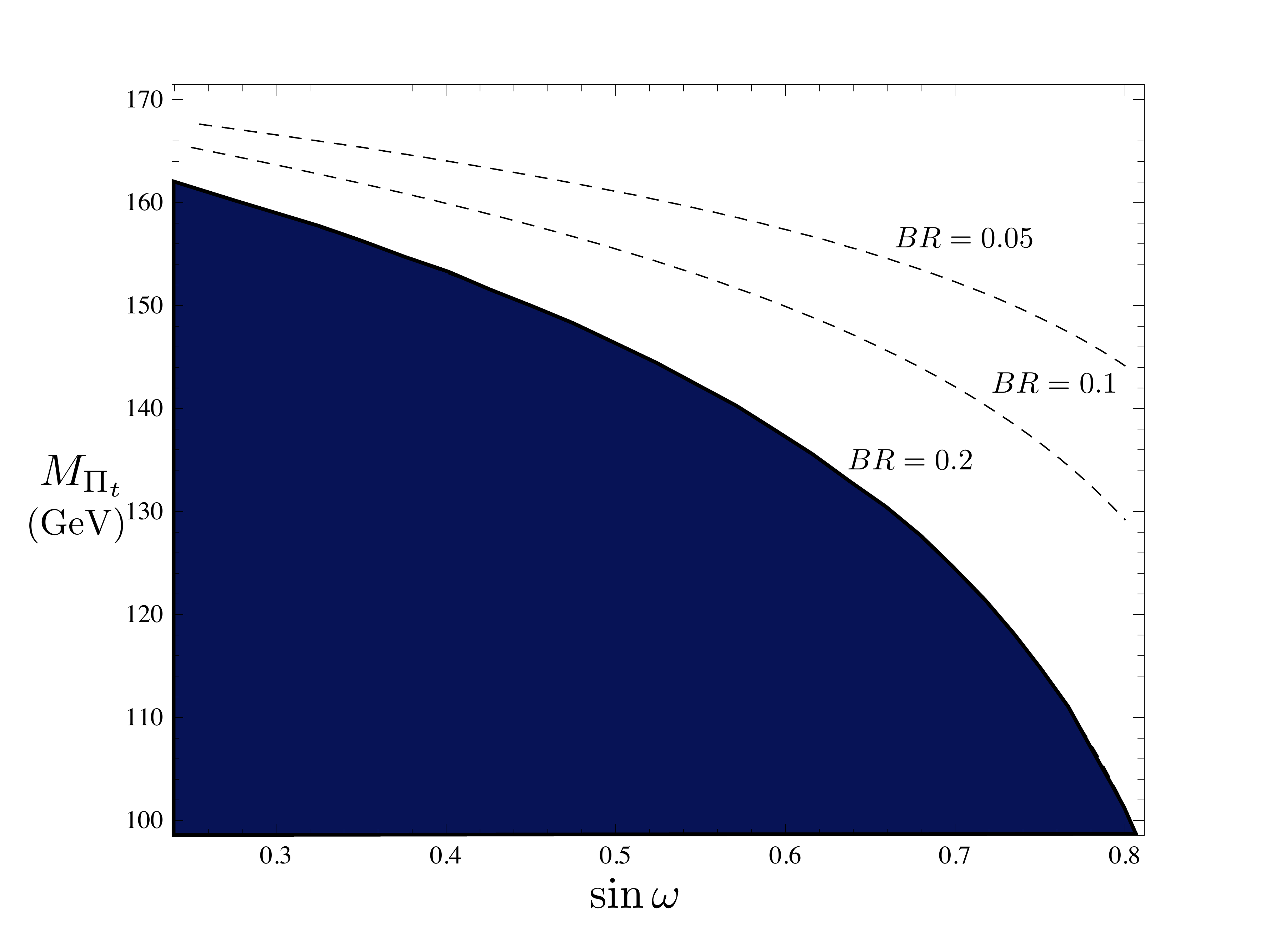}
\includegraphics[angle=0,width=3.5in]{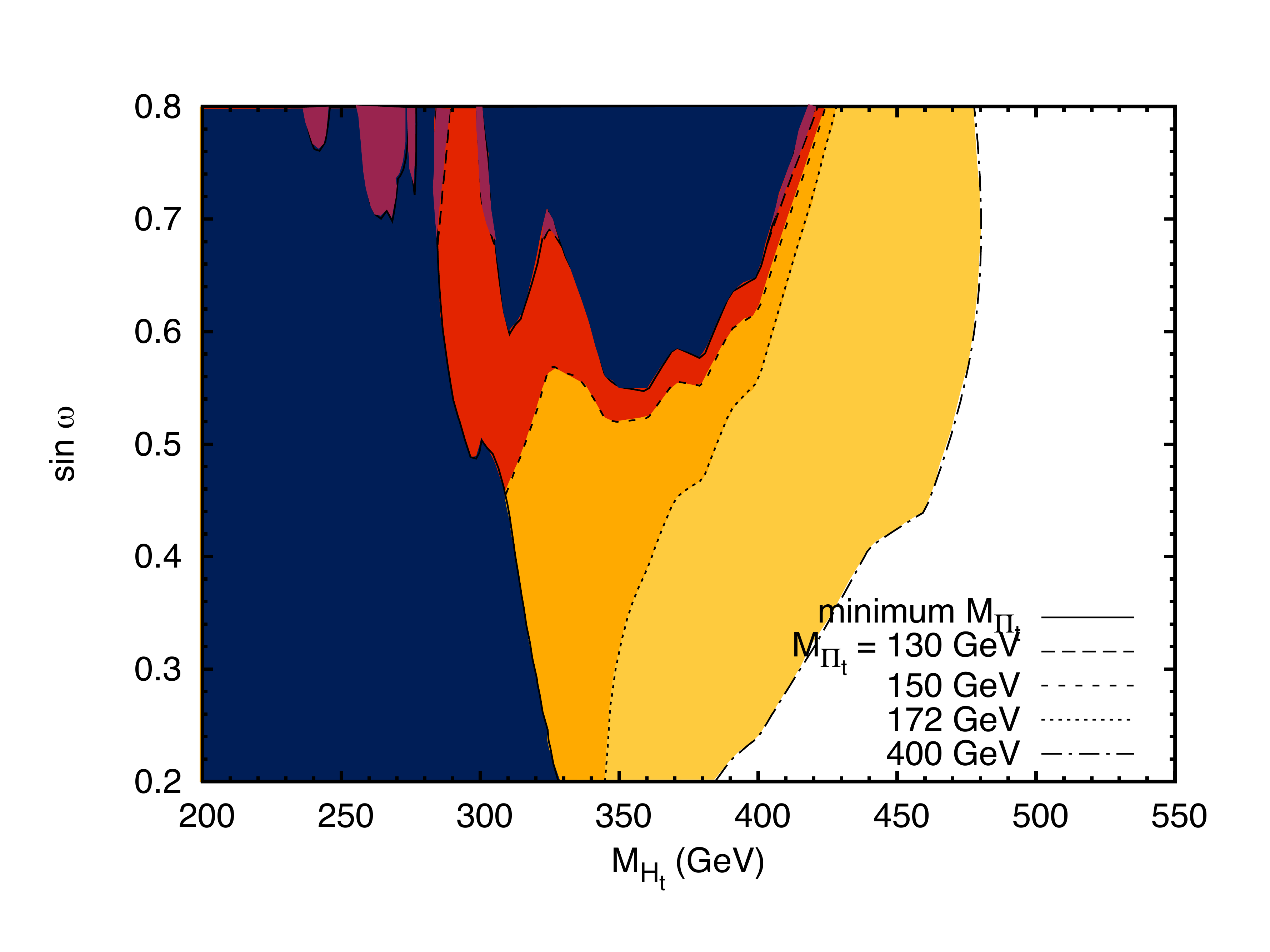}
\caption{Left: Contours of constant branching ratio $BR(t \to \Pi_t b)$, as calculated from eq. (\protect\ref{eq:tBR}) in 
the $(\sin\omega, M_{\Pi_t})$ plane, taking $m_t=172$ GeV and neglecting the bottom-quark mass. 
The dark blue region is excluded by Tevatron bound \protect\cite{Aaltonen:2009ke,Abazov:2009zh}, and $M_{\Pi_t}$ must 
lie above the $BR=0.2$ line for the corresponding value of $\sin\omega$. The
contours for $BR=0.1$ and 0.05 (dashed lines) are shown to indicate how this bound may evolve
in the future if the bound continues to improve.
Right: Regions in the $(M_{H_t}, \sin\omega)$ plane excluded by the   ATLAS 
\cite{ATLAS-higgs}  and CMS  \cite{CMS-higgs} upper bounds on
 $\sigma (pp \to H_t \to WW)$ for $M_{\Pi_t}= 130$ GeV (dark wine regions outside long-dashed lines), 150 GeV  (medium red regions above short-dashed line), 172 GeV (moderate orange region to left of dotted line) and 400 GeV (light gold region to left of dot-dashed line).   Very dark blue regions are excluded for top-pion masses that saturate
 the Tevatron bound for a given value of $\sin\omega$.}
\label{fig:contourplot}
\end{center}
\end{figure*}

At low energies, any top-condensate model includes the composite weak-doublet scalar boson with the 
same quantum numbers as the fundamental scalar introduced in the standard model \cite{Chivukula:1990bc}.
The vacuum expectation value of the composite weak-doublet scalar boson, $f_t$, combined with the technipion decay-constant
of the technicolor theory, $F$, yield the usual electroweak scale
\begin{equation}
v^2 = \frac{1}{\sqrt{2} G_F} = f^2_t + F^2\approx (246\, {\rm GeV})^2,
\label{weak-scale}
\end{equation}
where $G_F$ is the weak-interaction Fermi constant. Motivated by this relation, we define
an angle $\omega$ such that $f_t \equiv v \sin \omega$. 
Three of the degrees of freedom of the composite
scalar mix with the states in the technicolor spectrum which are the analogs of the pions of QCD. Through
the Higgs mechanism \cite{higgs-mechanism}, one set of linear combinations become the longitudinal states of the $W^\pm$ and $Z$. The orthogonal combinations, which we denote $\Pi^\pm_t$ and $\Pi^0_t$, remain in the spectrum
and are referred to as ``top-pions." Ignoring (small) electromagnetic corrections to their masses, 
the charged and neutral  top-pions are degenerate.
The fourth degree of freedom in the composite scalar\footnote{In principle,
this degree of freedom will also mix with a state in the technicolor spectrum, a state analogous to the
putative ``sigma" particle in QCD. In practice such a state has a mass of order a TeV or higher, and this
mixing is negligible.}, which we denote $H_t$, is the neutral ``top-Higgs."
The phenomenology of the scalar sector of top-condensate
models is determined by the masses of the top-Higgs and top-pions, $M_{H_t}$ and $M_{\Pi_t}$, and
the value of $\sin\omega$. We consider the range of allowed masses and mixing angles, and the
corresponding scalar phenomenology, below.

Quantitative analyses of the strong topcolor dynamics
\cite{Miransky:1988xi,Miransky:1989ds,Nambu:1989jt,Marciano:1989xd,Bardeen:1989ds} use the Nambu--Jona-Lasinio \cite{Nambu:1961tp} (NJL) approximation to the topcolor interactions, 
solved in the large-$N$ limit \cite{'tHooft:1973jz}. In this limit, we find the Pagels-Stokar
relation \cite{Pagels:1979hd} for $f_t$
\begin{equation}
f_{_t}^2 = \dfrac{N_c}{8\pi^2} m_{t,dyn}^2 \ln\left(\dfrac{\Lambda^2}{m_{t,dyn}^2}\right)~,
\label{eq:TC2-ft}
\end{equation}
where $\Lambda$ is
the cutoff of the effective NJL theory, which is expected to be of order a few to tens of TeV \cite{Braam:2007pm}, 
and $m_{t,dyn}$ denotes the portion of the top-quark mass arising from topcolor.
The portion of the top-quark mass arising from technicolor interactions 
(more properly, extended technicolor
\cite{Dimopoulos:1979es,Eichten:1979ah} interactions) is expected to be less than or
of order the bottom-quark mass, and hence $m_{t,dyn} \approx m_t$ \cite{Hill:1994hp, Hill:2002ap}.
Varying $\Lambda$ between 1 and 20 TeV, we find 
\begin{equation}
0.25 \laem \sin\omega \laem 0.5~.
\label{eq:sinomega}
\end{equation}

In the large-$N$/NJL approximation, we find $m_{H_t} = 2 m_{t,dyn} \approx 350$ GeV. This relation can
be modified via QCD interactions which, in the leading-log approximation (here $\log(\Lambda/M_{H_t})$)
tend to lower the top-Higgs mass \cite{Bardeen:1989ds,Hill:1985tg}. In addition, there can be additional 
(non leading-log) corrections coming from interactions in the topcolor theory that are not included in
the NJL approximation, and also corrections that are subleading in $N$. Therefore, while top-Higgs masses
less than or of order 350 GeV are generically expected in these theories, we will display results for masses
between 200 and 600 GeV.

The top-Higgs couples to top-quarks and to pairs of electroweak bosons, and it does so
in a characteristic manner. Since topcolor interactions give rise to $m_{t,dyn}$, the bulk of the top
mass, and since the expectation value of the composite weak scalar doublet is $f_t = v \sin\omega$,
the Yukawa coupling of $H_t$ to top-quarks is 
\begin{equation}
y_{H_t} = \frac{\sqrt{2} m_{t,dyn}}{f_t} \approx \frac{y_t}{\sin\omega}~,
\label{eq:yukawaht}
\end{equation}
where $y_t = \sqrt{2} m_t/v$ is the standard model top-quark Yukawa coupling. Hence, the top-Higgs couples
more strongly to top-quarks than does the standard model Higgs boson. This enhanced coupling implies
an enhancement for top-Higgs production via gluon fusion, relative to the analogous process
for the standard model Higgs boson \cite{Georgi:1977gs},
\begin{equation}
\frac{\sigma_{gg} (pp \to H_t)}{\sigma_{gg}(pp \to H_{SM})}
= \frac{\Gamma(H_t \to gg)}{\Gamma(H_{SM} \to gg)} \approx
\frac{1}{\sin^2\omega}~.
\label{eq:htgg}
\end{equation}
In contrast, since the bulk of electroweak symmetry breaking is provided by technicolor (see eq. (\ref{weak-scale})),
the coupling of the top-Higgs to  vector boson pairs 
is suppressed relative to the standard model
\begin{equation}
g_{H_t WW/ZZ} = \sin \omega \cdot g_{H_{SM} WW/ZZ}~.
\label{eq:wwht}
\end{equation}
Hence the top-Higgs vector-boson
fusion (VBF) production cross section, and the partial width of $H_t$ to vector boson
pairs, are also suppressed
\begin{equation}
\frac{\sigma_{VBF} (pp \to H_t)}{\sigma_{VBF}(pp \to H_{SM})}
= \frac{\Gamma(H_t \to W^+W^-/ZZ)}{\Gamma(H_{SM} \to W^+W^-/ZZ)}
 \approx
\sin^2\omega~.
\label{eq:htWW}
\end{equation}
Given that the dominant top-Higgs production pathway is enhanced (see eq. (\ref{eq:htgg})), the suppression of the vector boson fusion pathway is not a major concern.

The crucial issue for the LHC phenomenology
of the top-Higgs is the branching ratio $BR(H_t \to WW/ZZ$): if this branching ratio is sufficiently large, the ATLAS
\cite{ATLAS-higgs} and CMS \cite{CMS-higgs} detectors will be sensitive to the existence of a top-Higgs.
As we will now discuss, the branching ratio of the top-Higgs to vector bosons, in turn, depends on the mass of the top-pion.
Unlike the top-Higgs mass, the top-pion masses depend on the amount of top-quark mass arising
from the (extended) technicolor sector, and on the effects of electroweak gauge interactions
\cite{Hill:1994hp,Hill:2002ap}. These masses are therefore more model-dependent. Since
top-pions are in the electroweak symmetry breaking sector, we expect them to be lighter than
a TeV.

Top-pions cannot be too light, however. If the charged top-pion $\Pi_t^+$ were 
lighter than the top quark, it would potentially appear in top decays, $t \to \Pi_t^+ b$.  The Tevatron experiments 
have searched for this process in the context of two-Higgs-doublet models and set upper bounds of about 10--20\% 
on the branching fraction of $t \to H^+ b$, with $H^+$ decaying to $\tau \nu$ or $c \bar{s}$ (actually,  two jets) 
\cite{Aaltonen:2009ke,Abazov:2009zh}, as the top-pion would also do. The branching ratio
$BR(t \to \Pi^+_t b)$  \cite{Chivukula:2011ag}, is\footnote{These expressions were derived in the ``top-triangle moose" model \protect\cite{Chivukula:2009ck,Chivukula:2011ag}, a low-energy effective theory for TC2, neglecting  corrections
due to heavy particles (Dirac fermions and extra vector-bosons) that are present in the top-triangle model, and focusing on
 the generic TC2 couplings. 
The top-triangle model-dependent corrections from the heavy states are of order a few percent, and their inclusion here or in our other computations 
would not change the results. The insensitivity of our analysis to model-dependent top-triangle effects is a confirmation that our
results are generic for TC2 and other top-condensate models whose spectra have only $H_t$ and $\Pi_t$ particles present at low-energies  in
the top-mass generating sector of the theory.}
\begin{widetext}
\begin{equation}
BR(t\to \Pi^+_t b)  \approx \frac{\Gamma^{TC2}(t\to \Pi^+_t b)}{\Gamma^{SM}(t\to W^+b) + \Gamma^{TC2}(t\to \Pi^+_t b)} \label{eq:tBR}
 = \frac{\cot^2\omega\left(1-\frac{M^2_{\Pi_t}}{m^2_t}\right)^2}{\left( 1 + \frac{2 M_W^2}{m_t^2} \right)
	\left( 1 - \frac{M_W^2}{m_t^2} \right)^2+ \cot^2\omega\left(1-\frac{M^2_{\Pi_t}}{m^2_t}\right)^2} ~,
\end{equation}
\end{widetext}
where we neglect the bottom-quark mass.
From this we see that, for a given value of $\sin\omega$, there is a minimum value of $M_{\Pi_t}$ such
that $BR(t \to \Pi^+ t) \lesssim 0.2$. This lower bound on $M_{\Pi_t}$ is illustrated in the left panel of Fig. \ref{fig:contourplot}.

We may now return to the value of $BR(H_t \to WW/ZZ)$, which is crucial for understanding the
LHC limits on these models. If kinematically allowed, the top-Higgs will decay predominantly to $\Pi_t + W/Z$,
$2\Pi_t$, or $t\bar{t}$.\footnote{In this analysis, we neglect off-shell decays since the three-point couplings 
for these processes 
are the same order of magnitude. Adding these processes would not change our results.} The relevant couplings may be found in \cite{Chivukula:2011ag}, where we ignore the
small model-dependent corrections arising from heavy particles (see footnote 3).
For decays to top-pion plus gauge boson, 
\begin{eqnarray}
  \Gamma(H_t \to \Pi^{\pm}_t W^{\mp}) &=& \frac{\cos^2\omega}{8 \pi v^2}
  M_{H_t}^3 \beta_W^3, \nonumber \\
  \Gamma(H_t \to \Pi^0_t Z) &=& \frac{\cos^2\omega}{16 \pi v^2} 
  M_{H_t}^3 \beta_Z^3,
  \label{eq:HPiV}
\end{eqnarray}
where
\begin{equation}
  \beta_V^2 \equiv \left[ 1 - \frac{(M_{\Pi_t} + M_V)^2}{M_{H_t}^2} \right]
  \left[ 1 - \frac{(M_{\Pi_t} - M_V)^2}{M_{H_t}^2} \right].
\end{equation}
For decays to two top-pions,
\begin{equation}
  \Gamma(H_t \to \Pi_t^+ \Pi_t^-) =
 2 \Gamma(H_t \to \Pi_t^0 \Pi_t^0) =
  \frac{\lambda^2_{H\Pi\Pi}}{16 \pi M_{H_t}} 
  \sqrt{1 - \frac{4 M_{\Pi_t}^2}{M_{H_t}^2}},
\end{equation}
where 
\begin{equation}
  \lambda_{H\Pi\Pi} = \frac{1}{2 v \sin\omega}
  \left[ \left( M_{H_t}^2 - 2 M_{\Pi_t}^2 \right) \cos 2\omega
    + M_{H_t}^2 \right].
\end{equation}
And for decays to top-quark pairs,
\begin{equation}
\Gamma(H_t \to t\bar{t}) = \frac{3  m^2_t}{8\pi v^2 \sin^2\omega} M_{H_t}
\left(1-\frac{4 m^2_t}{M^2_{H_t}}\right)^{3/2}~.
\label{eq:Httbar}
\end{equation}
By comparison, as we have previously discussed, the width to gauge-bosons is suppressed
by $\sin^2\omega$
\begin{eqnarray}
	\Gamma(H_t \to W^+W^-) &=&
	\frac{M_{H_t}^3 \sin^2\omega}{16 \pi v^2}
	\sqrt{1 - x_W} \left[ 1 - x_W + \frac{3}{4} x_W^2 \right],
	\nonumber \\
	\Gamma(H_t \to ZZ) &=&
	\frac{M_{H_t}^3 \sin^2\omega}{32 \pi v^2}
	\sqrt{1 - x_Z} \left[ 1 - x_Z + \frac{3}{4} x_Z^2 \right],
\label{eq:HVV}
\end{eqnarray}
where $x_V = 4 M_V^2 / M_{H_t}^2$.

If all decays are kinematically unsuppressed, and for the mass ranges we consider,
a hierarchy of decay widths emerges:
\begin{equation}
\Gamma(H_t\to 2\Pi_t)\ \gtrsim\  \Gamma(H_t \to t\bar{t}),\ \Gamma(H_t \to \Pi_t + W/Z) \ \gtrsim\ \Gamma(H_t \to WW/ZZ)~.
\label{eq:order}
\end{equation}
In particular, if kinematically allowed, the top-Higgs decays  predominantly into pairs of top-quarks or top-pions. As we shall
see, this implies that LHC searches are particularly sensitive to top-Higgs masses less than about 400 GeV. For this range
of top-Higgs masses, LHC sensitivity depends crucially on the top-pion masses and whether the top-Higgs decays to either
a top-pion pair or top-pion plus vector boson is allowed.  LHC searches are most sensitive
when these decay modes are suppressed and the $BR(H_t \to WW/ZZ)$ is therefore as large as possible.

\section{LHC Limits on the Top-Higgs}

\begin{figure*}[t]
\begin{center}
\includegraphics[angle=0,width=3.5in]{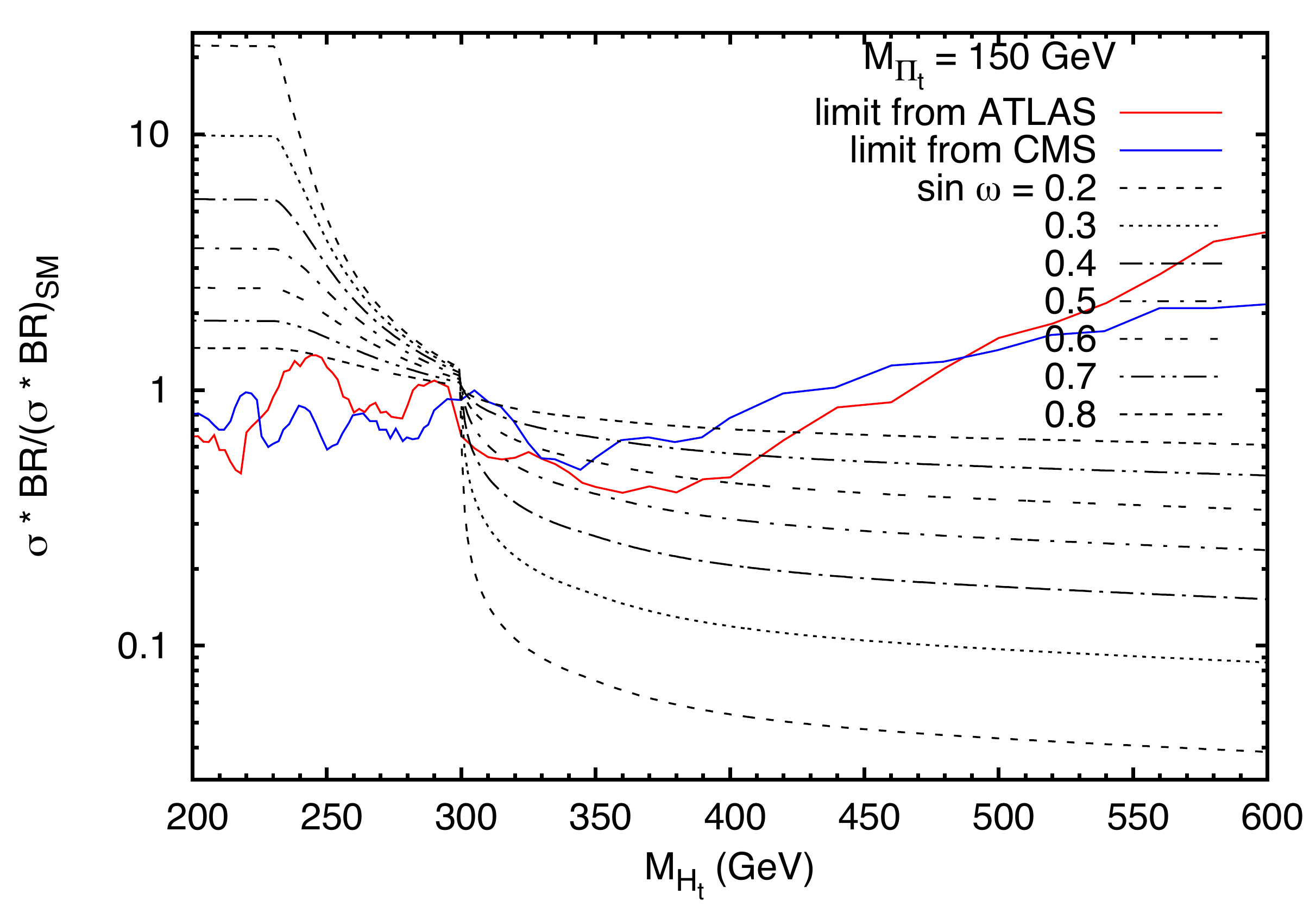}
\includegraphics[angle=0,width=3.5in]{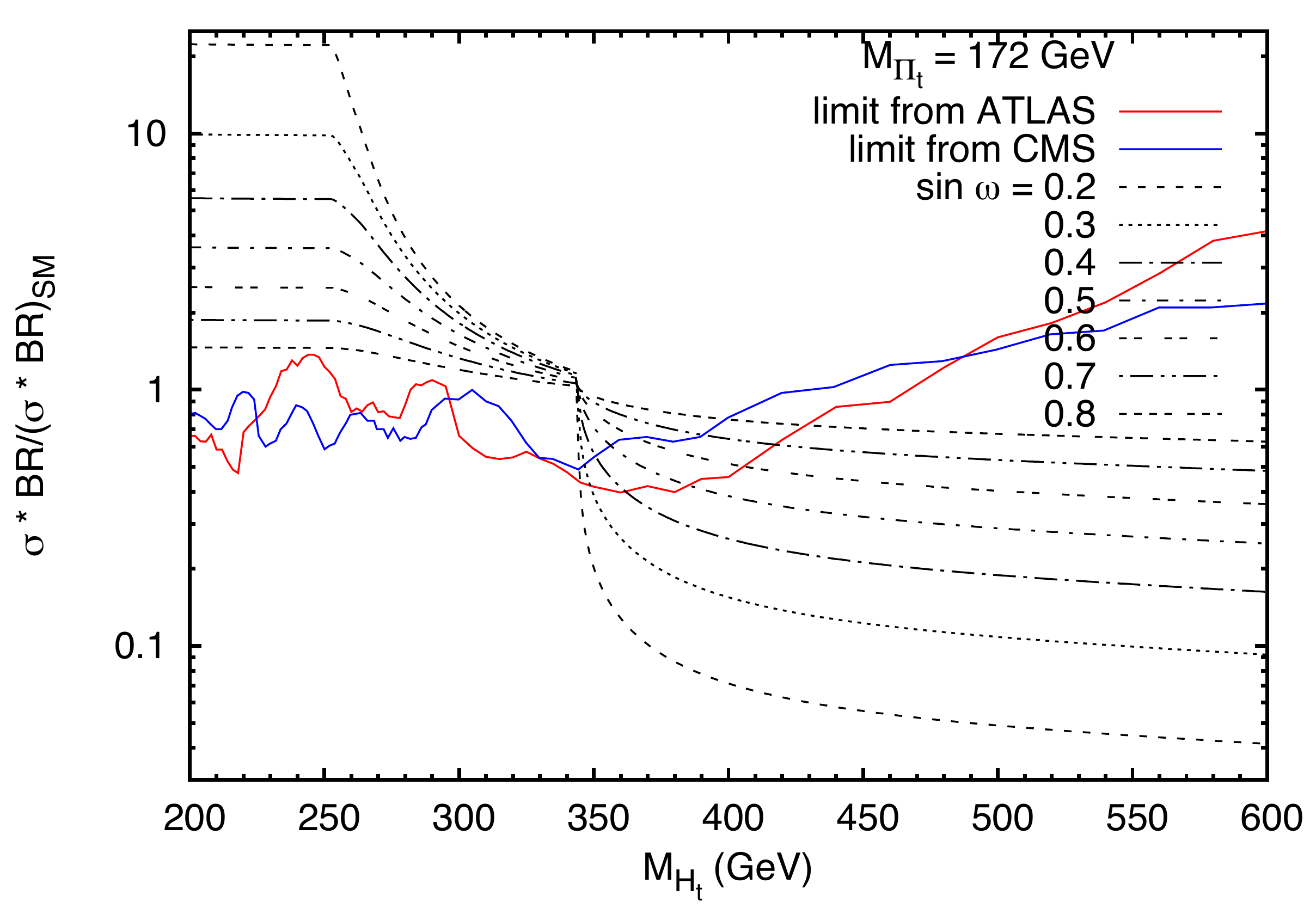}\\
\includegraphics[angle=0,width=3.5in]{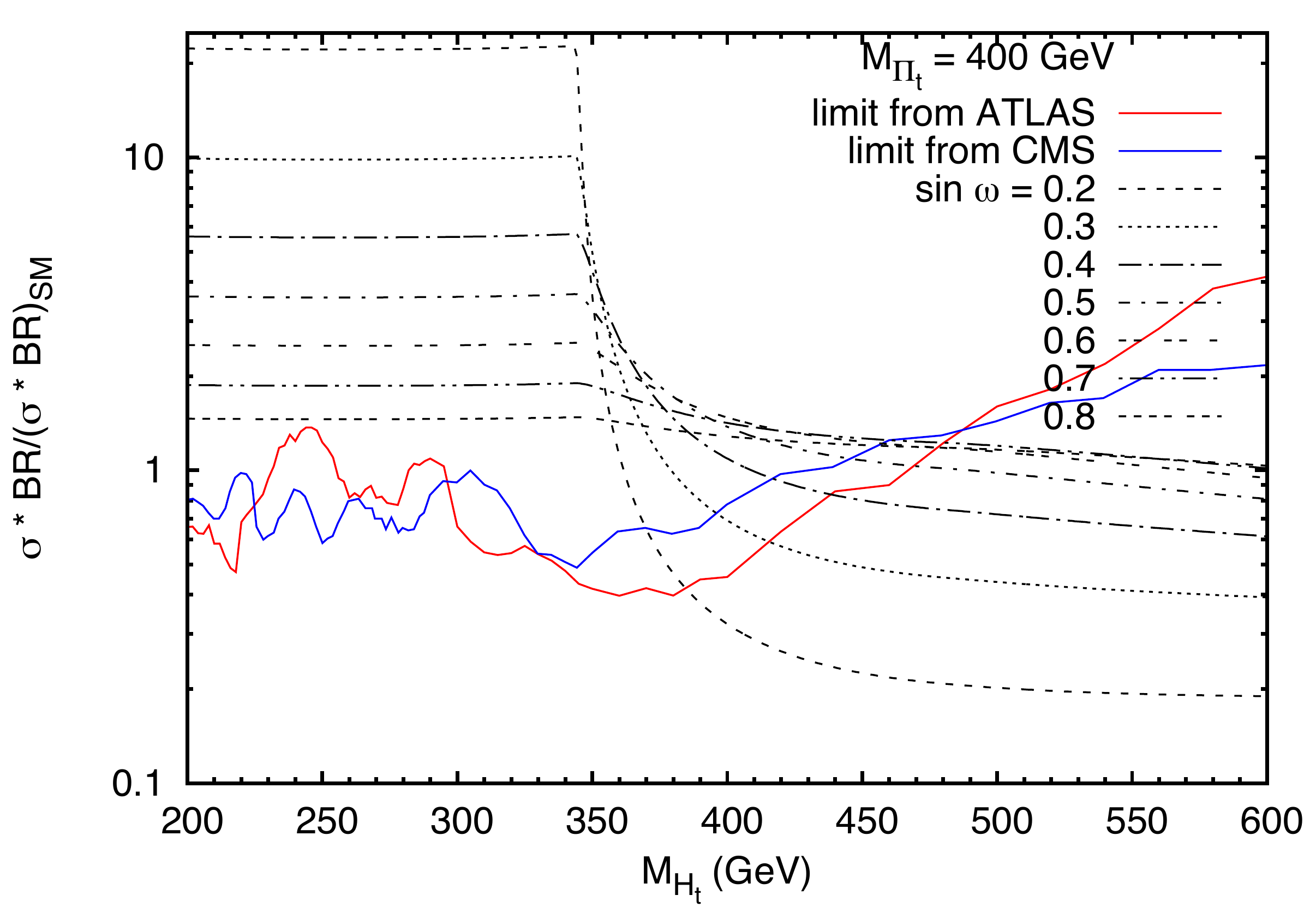}
\includegraphics[angle=0,width=3.5in]{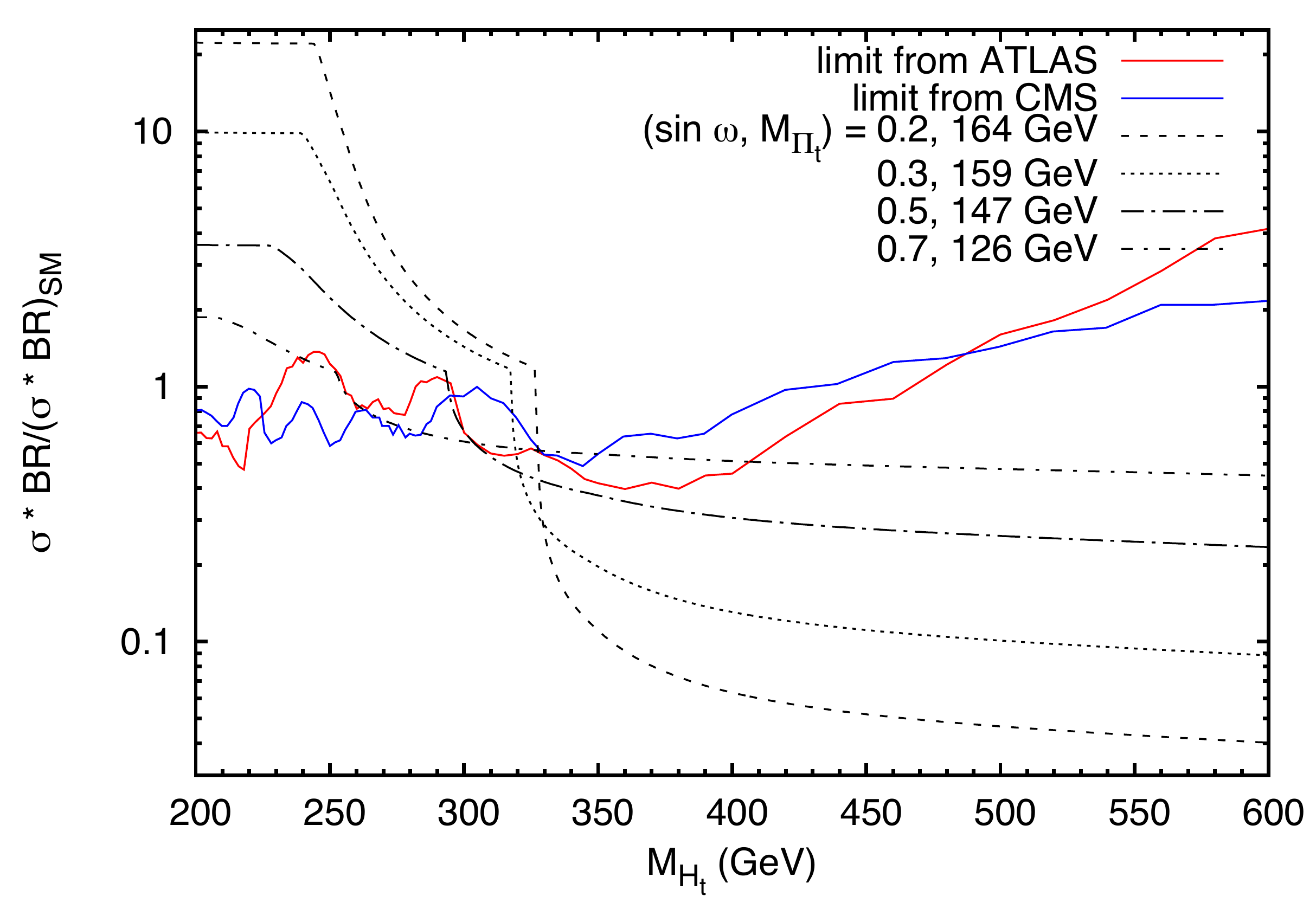}
\caption{Upper Left: LHC top-Higgs production cross section times $WW$ branching ratio,
$\sigma (pp \to H_t \to WW)$, relative to that of the standard model as a function of top-Higgs mass,
for a top-pion mass $M_{\Pi_t} = 150$ GeV and
various values of $\sin\omega = f_t/v$. Also shown are the corresponding ATLAS \cite{ATLAS-higgs} and CMS \cite{CMS-higgs}  95\% CL upper bounds on this ratio. Note the sharp drop in the branching ratio
when the $\Pi_t W/Z$ mode opens, at approximately $M_{H_t}=230$ GeV, and the further drop when
the $2\Pi_t$ mode opens, at approximately 300 GeV. Upper Right: Same plot, $M_{\Pi_t}=172$ GeV. Here, since $M_{\Pi_t} \simeq m_t$, the sharp
drop in the branching ratio occurs when $t\bar{t}$ and $2\Pi_t$ modes both open near 350 GeV.  Lower Left: Same plot, $M_{\Pi_t}=400$ GeV. Here, again, the branching
ratio falls sharply above 350 GeV as the $t\bar{t}$ decay mode opens. Regions excluded by these plots are shaded in medium hues of red and gold in the
right panel of Fig. \protect\ref{fig:contourplot}.
Lower Right: Same plot, for combinations of $\sin\omega$ and top-pion mass $M_{\Pi_t}$ that saturate the Tevatron
bound on $BR(t \to \Pi b)\lesssim 0.2$ \cite{Aaltonen:2009ke,Abazov:2009zh}. The regions excluded by this plot are shaded very dark blue in
the right panel of  Fig. \protect\ref{fig:contourplot}.}
\label{fig:rateplot}
\end{center}
\end{figure*}

We now turn to the LHC limits on the top-Higgs that follow from the recent
ATLAS \cite{ATLAS-higgs} and CMS \cite{CMS-higgs} searches for the standard model
Higgs boson. For the reasons described above, we consider top-Higgs masses ranging from 200 to
600 GeV. In this mass range, the standard model Higgs boson
is produced primarily through gluon fusion and secondarily through
vector-boson fusion \cite{new}. The strongest limits \cite{ATLAS-higgs, CMS-higgs} in this mass range come from
searching for the Higgs boson decaying to $W^+W^-$ or $ZZ$. 
In the narrow-width approximation for $H_t$,\footnote{We will justify the validity of this approximation for the
regions in which the ATLAS and CMS bounds apply.}  the inclusive cross section $\sigma(pp \to H_t \to WW/ZZ)$ 
may be related to the corresponding standard model cross section through the expression
\begin{widetext}
\begin{align}
\frac{\sigma(pp \to H_t \to WW/ZZ)}{\sigma(pp \to H_{SM} \to WW/ZZ)}
& = \frac{\left[ \sigma_{gg}(pp \to H_t) + \sigma_{VBF}(pp \to H_t)\right] BR(H_t \to WW/ZZ)}
{\left[ \sigma_{gg}(pp \to H_{SM}) + \sigma_{VBF}(pp \to H_{SM})\right] BR(H_{SM} \to WW/ZZ)}
\label{eq:BRscaling}\\
& \approx
\frac{
\left(\frac{1}{\sin^2\omega} \sigma_{gg}(pp \to H_{SM}) +
\sin^2\omega \cdot\sigma_{VBF}(pp \to H_{SM})\right)}
{\sigma_{gg}(pp \to H_{SM}) + \sigma_{VBF}(pp \to H_{SM})}
\cdot \frac{BR(H_t \to WW/ZZ)}{BR(H_{SM} \to WW/ZZ)}~. \nonumber
\end{align}
\end{widetext}
While this relationship is appropriate for the ratio of  inclusive cross sections, the 
experimental limits  include detector-dependent effects
such as acceptances and efficiencies. To the extent that gluon-fusion and vector-boson fusion Higgs (or
top-Higgs) events differ, then this equation is only approximately correct. For Higgs masses between 200 and 600 GeV, however, the vector-boson fusion cross section accounts for only ${\cal O}(10\%)$ of
the standard model Higgs production cross-section, and we therefore expect the 
scaling relation will hold to better than this level of accuracy.  
We compute BR($H_t \to WW/ZZ$) using eqs. (\ref{eq:HPiV}) - (\ref{eq:HVV}), 
and $BR(H_{SM} \to WW/ZZ)$ using eqs. (\ref{eq:Httbar}) - (\ref{eq:HVV}) with $\sin\omega \to 1$,
and  we obtain the 7 TeV LHC standard model production cross sections $\sigma_{gg,VBF}(pp \to H_{SM})$ 
from \cite{new}. Putting this all together, we use eq. (\ref{eq:BRscaling}) to convert the limits on the standard model Higgs in \cite{ATLAS-higgs, CMS-higgs} into
limits on the top-Higgs in TC2 models. 

In Fig. \ref{fig:rateplot} we show the ratio of $\sigma(pp \to H_t \to WW/ZZ)$ divided by the corresponding 
quantity for the standard model Higgs, as a function of $M_{H_t}$ for various
values of $\sin\omega$, and for $M_{\Pi_t} = 150$ GeV (upper left), 172 GeV (upper right) and 400 GeV (lower left). Also plotted on these
graphs are the recent 95\% CL LHC upper bounds \cite{ATLAS-higgs, CMS-higgs} on these quantities. 
For $M_{\Pi_t}=150$ GeV, note the sharp drop in the branching ratio
when the $\Pi_t W/Z$ mode opens, at approximately $M_{H_t}=230$ GeV, and the further drop when
the $2\Pi_t$ mode opens, at approximately 300 GeV. Because of these drops in the branching ratios for vector boson pairs, the LHC limits
on the top-Higgs are weaker when the top-pions are lighter. For $M_{\Pi_t} \simeq m_t = 172$ GeV, the sharp
drop in the branching ratio occurs when $t\bar{t}$ and $2\Pi_t$ open near 350 GeV. 
Finally, for $M_{\Pi_t}=400$ GeV, again the branching ratio falls sharply above 350 GeV as the $t\bar{t}$ decay mode 
opens. Since $BR(H_t \to WW/ZZ)$ is larger in the regions where $M_{H_t} < M_{\Pi_t} + M_{W,Z}$, the LHC
limits on the top-Higgs are substantially stronger for heavier $M_{\Pi_t}$.  The regions excluded by these plots in the $(M_{H_t}, \sin\omega)$ plane
are shaded in hues of red, orange, and gold in the right panel of Fig. \protect\ref{fig:contourplot}.

From the left panel of Fig. \ref{fig:contourplot}, we see that the minimum $M_{\Pi_t}$ that satisfies the Tevatron upper bound
on $BR(t \to \Pi^+ b)$ \cite{Aaltonen:2009ke,Abazov:2009zh} depends on $\sin\omega$. 
In the lower-right panel of Fig. \ref{fig:rateplot} we plot the LHC top-Higgs production cross section times $WW$ branching ratio,
$\sigma (pp \to H_t \to WW)$, relative to that of the standard model as a function of top-Higgs mass,
for combinations of $\sin\omega$ and top-pion mass $M_{\Pi_t}$ that saturate the Tevatron
bound on $BR(t \to \Pi^+ b)$ \cite{Aaltonen:2009ke,Abazov:2009zh}.  We also show
 the corresponding ATLAS \protect\cite{ATLAS-higgs} and CMS \cite{CMS-higgs}  95\% CL upper bounds on this ratio. 
Again, note the  drop in the branching ratio when the $\Pi_t W/Z$ mode opens. The regions 
excluded by this plot are shaded very dark blue in the right panel of Fig. \protect\ref{fig:contourplot}. 

In translating the ATLAS and CMS limits on the standard model Higgs boson to the top-Higgs,
we have used the narrow-width approximation. This breaks down for
sufficiently large $M_{H_t}$ and small $\sin\omega$. However, as we have seen, in the region 
where the ATLAS and CMS bounds apply to the top-Higgs, the decays to $WW/ZZ$ dominate and those to $\Pi_t W/Z$ or $2\Pi_t$ are kinematically suppressed. For these parameter values the width of the top-Higgs is 
comparable to that of the standard model Higgs, and hence our scaling is valid.

Fig. \ref{fig:contourplot} summarizes our results for bounds on the top-Higgs in models with strong top
dynamics. The regions shaded very dark blue are completely excluded: on the left, directly from
the Tevatron bounds and on the right from the LHC searches for the standard model Higgs boson. We see that
top-Higgs masses of 300 GeV or less are excluded at 95\% CL for any value of $\sin\omega$ and for
$M_{\Pi_t^+} \gtrsim 150$~GeV.  Moreover, when the top-pion is heavier than the top quark, 
all of the generic parameter range in TC2 models ($ 0.25 \laem \sin\omega \laem 0.5$ and $M_{H_t} \lesssim 2 m_t$) 
is excluded at 95\% CL.

\section{Discussion}

In this paper we have used the LHC limits on the standard model Higgs boson
\cite{ATLAS-higgs, CMS-higgs} to constrain the top-Higgs state predicted in
many models with new dynamics that couples strongly to top quarks, including topcolor assisted technicolor, top seesaw, and certain Higgsless models. Such a state
generically couples strongly to top-quarks, and is therefore produced through gluon
fusion at an enhanced rate relative to the standard model Higgs boson. If the spectrum
of the theory allows the branching ratio of the top-Higgs to vector boson pairs
to be sufficiently high, which happens if the corresponding top-pion is sufficiently heavy, then current
LHC searches for the standard model Higgs boson exclude the existence of the
top-Higgs, as summarized in Fig. \ref{fig:contourplot}.  

Our results show that the relatively light top-Higgs states expected in generic TC2 models are tightly constrained.  Moreover, as described in footnote 3, we have used the effective theory supplied by the top triangle moose to confirm that these conclusions apply broadly to top-condensate models that have only $H_t$ and $\Pi_t$ particles in the low-energy spectrum of the sector responsible for generating the top quark mass.

However, models with top-Higgs
masses larger than 350 GeV are still allowed.  In this region, for small $\sin\omega$, the top-Higgs becomes
a very broad state decaying predominantly into top-quark or top-pion pairs and LHC searches for this state may
be difficult. Within the context of a TC2 model, 
however, it would be difficult to reach that region of parameter space.  In principle, the non leading-log (or sub-leading in $1/N$) corrections to the NJL
approximation to the topcolor interactions could shift the top-Higgs mass toward substantially larger values. However, precision flavor and electroweak analyses \cite{Braam:2007pm}
prefer larger values of the cutoff $\Lambda$ and make it unlikely that these effects are large enough to do so. 

One avenue to constructing a viable dynamical theory with large $M_{H_t}$ might be to pair technicolor with a ``top-seesaw" \cite{Dobrescu:1997nm,Chivukula:1998wd} sector rather than a topcolor sector \cite{Fukano:2011fp}.  
In a top-seesaw model, condensation of a heavy seesaw top-partner fermion breaks the electroweak symmetry,  thereby severing the link between
the top-quark mass and $f_t$  illustrated by eq. (\ref{eq:TC2-ft}). Because of the increased value of $f_t$, a seesaw-assisted technicolor theory would feature both larger values of $M_{H_t}$ and higher values of $\sin\omega$ than are typical for TC2.
In fact, just as the top-triangle moose serves as a low-energy effective theory for the top-Higgs and top-pion sectors of 
TC2 in the region of moose parameter space where $f_t$ is relatively small, it may also be
viewed as  a low-energy effective theory for top-seesaw assisted technicolor when $f_t$ is relatively large.    In essence,  a top-seesaw
assisted technicolor theory smoothly interpolates between TC2 and the standard model with a heavy Higgs boson -- a situation that is potentially allowed in the presence of weak isospin violation
\cite{Chivukula:2000px,He:2001fz}.   

As additional LHC data is accumulated in the coming months, we anticipate that further searches for signs of a Higgs decaying to vector boson pairs will either reveal the presence of a top-Higgs or raise the lower bound on its mass.  In either case, the implications for theories with new strong top dynamics will be profound.

\bigskip

\begin{acknowledgments}
The authors thank Bogdan Dobrescu and Chris Hill for useful conversations.
BC and HEL were supported by the Natural Sciences and Engineering Research Council of Canada.  RSC and EHS were supported, in part, by the US National Science Foundation under grant PHY-0854889. They also gratefully acknowledge the hospitality of the Aspen Center for Physics, which
is supported in part by the National Science Foundation under Grant No. 1066293.
AM is supported by Fermilab operated by Fermi Research Alliance, LLC under contract number DE-AC02-07CH11359 with the US Department of Energy.

\end{acknowledgments}


\end{document}